\documentclass[useAMS,usenatbib,usegraphicx]{mn2e}

\newcommand{\kev}{keV}
\newcommand{\ergcms}{erg~cm~s$^{-1}$}
\newcommand{\fe}{Fe~K$\alpha$}
\newcommand{\etal}{et al.}
\newcommand{\mcg}{MCG--6-30-15}
\newcommand{\oeight}{O~\textsc{viii}}
\newcommand{\osev}{O~\textsc{vii}}
\newcommand{\nsev}{N~\textsc{vii}}
\newcommand{\nsix}{N~\textsc{vi}}
\newcommand{\csix}{C~\textsc{vi}}

\title[Soft X-ray lines from photoionized accretion discs]
  {Soft X-ray emission lines from photoionized accretion discs:
  constraints on their strength and width}
\author[D.\ R.\ Ballantyne, R.\ R.\ Ross \& A.\ C.\ Fabian]
  {D.~R.~Ballantyne$^1$\thanks{drb@ast.cam.ac.uk}, R.~R.~Ross$^2$ and
  A.~C.~Fabian$^1$\\
  $^1$Institute of Astronomy, Madingley Road, Cambridge CB3 0HA \\
  $^2$Physics Department, College of the Holy Cross, Worcester, MA
  01610, USA}

\pagerange{\pageref{firstpage}--\pageref{lastpage}}
\pubyear{2002}

\usepackage{times}

\begin{document}

\label{firstpage}

\maketitle

\begin{abstract}
We consider the properties of soft X-ray emission lines in the
reprocessed emission from a photoionized accretion disc. Observations
of these lines will be important in determining the ionization state
and metallicity of the innermost regions of active
galaxies. Calculations of reflection from a constant-density disc with
an ionization parameter, $\xi$, between 250 and 1000~\ergcms\ show
that emission from \oeight\ Ly$\alpha$ will dominate the soft X-ray
emission spectrum. There is also significant emission from \csix,
\nsev, \osev, as well as Fe~\textsc{xvii}--\textsc{xix}. As $\xi$ is
increased these lines become weaker and are broadened by Compton
scattering. Significantly increasing the O abundance primarily
strengthens the \osev\ line, making it as large or larger than the
\oeight\ line. The nitrogen and carbon lines are quite weak with
equivalent widths (EWs) $<$30~eV, even with an increase in the N
abundance. A hydrostatic ionized disc model has a more realistic
density structure and shows a similar spectrum, but with the lines
weaker and broader. This is a result of the hot ionized skin at the
surface of the disc. We apply these results to the controversial claim
of soft X-ray relativistic lines in the \textit{XMM-Newton} spectrum
of \mcg. We are unable to find a situation where \oeight\ has the
required EW without substantial emission from \osev. Furthermore,
Compton scattering results in the blue wing of the \oeight\ line to be
much broader than the $\ll 10$~eV drop observed in the data. We
conclude that soft X-ray accretion disc lines will, in general, be
weak and broad features and are unlikely to produce sharp edges in the
data.
\end{abstract}


\section{Introduction}
\label{sect:intro}
Features observed in the hard X-ray spectra of active
galactic nuclei (AGN) are one of the most direct probes of the
structure of accretion discs close to the central black hole. The
prominent iron K$\alpha$ emission line at 6.4~\kev\ and Compton hump
at $\sim$20--30~\kev\ in the spectra of many Seyfert~1 galaxies
indicates that optically thick material is being illuminated by the
hard X-ray source \citep{pou90}. As pointed out by \citet{fab89}, if
this reflected emission arose from the accretion disc, then
relativistic and Doppler effects would broaden the emission lines in a
characteristic way. Observations of such a line allows a determination
of not only the accretion disc structure close to the black hole, but
also the space-time geometry in the strong gravity regime. The
\textit{ASCA} observatory (1993--2001) finally provided the
sensitivity to search for relativistically broadened \fe\ lines, and
detected a clear example in the spectrum of the bright Seyfert~1
galaxy \mcg\ \citep{tan95}. Since that initial discovery, broad \fe\
lines have been found in many other Seyferts \citep{nan97,yaq02}, and
are now an integral part of AGN phenomenology \citep{fab00}.

Calculations of the reprocessed emission from an irradiated slab of
gas have shown that the \fe\ line, by virtue of its large fluorescent
yield and relatively high cosmic abundance, is the strongest line in
the reflection spectrum \citep{gf91,mpp91}. Yet, strong lines from
abundant elements such as carbon, nitrogen and oxygen are also present
at lower energies, and, if detected, would provide further information
on the ionization state of the accretion disc. Models of reflection
from photoionized discs have shown that the soft X-ray features are
very sensitive to the ionization parameter of the disc material
\citep*{ros93,ros99}. Determining the ionization state of the disc
would constrain the magnitude of the illuminating radiation, and would
therefore place limits on the amount of accretion energy that is
released in the corona.

Although the soft X-ray spectra of Seyfert galaxies are complicated by
features such as warm absorbers and soft excesses, the sensitivity and
resolution provided by the instruments on board \textit{XMM-Newton}
and \textit{Chandra} could allow relativistic soft X-ray emission
lines to be found amongst the clutter and noise. Indeed, claims of
such lines from hydrogenic carbon, nitrogen and oxygen have already
been made based on \textit{XMM-Newton} Reflection Grating Spectrometer
(RGS) spectra of \mcg\ and Mrk~766 \citep{br01,sak02}. Clearly, the
time is right for a closer look at the properties of the soft X-ray
accretion disc lines.

Here, we employ detailed models of photoionized accretion discs to
investigate the properties (specifically, the strength and width) of
the C, N \& O lines found in the reflection spectra. A range of
ionization parameters are considered, and both constant density slabs
and hydrostatic atmospheres are used for the disc structure. The
calculations and assumptions are outlined in the following section,
with the results presented in Section~\ref{sect:res}. Finally, we
discuss the implications of our results in light of the recent
\textit{XMM-Newton} observations and draw conclusions in
Section~\ref{sect:discuss}.     

\section{Computations}
\label{sect:comp}
Models of illuminated accretion discs have greatly increased in
sophistication over the last couple of years. Initially, the
calculations had assumed that the surface of the disc was a constant
density slab \citep{ros93,zy94,mz95,ros99}, and so the resulting gas
structure and reflection spectrum could be characterized by the
ionization parameter
\begin{equation}
\xi= {4 \pi F_{\mathrm{X}} \over n_{\mathrm{H}}},
\label{eq:xi}
\end{equation}
where $F_{\mathrm{X}}$ is the X-ray flux incident on the slab, and
$n_{\mathrm{H}}$ is the hydrogen number density of the material. While
such a simple structure has some theoretical justification (a standard
optically-thick disc has a roughly constant vertical density profile in
the radiation pressure dominated limit; \citealt{ss73}), the heated
surface of the slab was out of pressure equilibrium with the remainder of the
material. Recently, more realistic calculations were performed with
the surface of the disc required to remain in hydrostatic balance
\citep*{nkk00,brf01}. Under this condition, the irradiated gas takes
on a more `two-phase' structure with an outer hot ($\approx$ Compton
temperature), ionized layer in pressure balance with a denser, colder
and more neutral inner zone. The spectral features in the reflection
spectra depend on the sharpness of the transition between the two layers
\citep{nkk00,br02}, but qualitatively appear to be a diluted
version of the constant density models. 

In this section, we detail ionized disc calculations made with both
the constant density code of \citet{ros93} and the hydrostatic code of
\citet{brf01}. Since these latter models are dependent on the unknown
accretion disc parameters, this paper concentrates on the results from
the constant density models. However, an example hydrostatic calculation
will be shown to illustrate the similarities and differences between
the two approaches. 

Important updates were made to both codes in order to better determine
the soft X-ray emission spectrum.  First, more elements were treated,
and more recombined ions of each element were included.  This was
necessary so that nitrogen lines would be included in the output
spectrum and so that the opacity would be treated accurately
throughout the soft X-ray portion of the spectrum.  In addition to
fully-ionized species, the following ions are included in the
calculations: C~{\sc iii--vi}, N~{\sc iii--vii}, O~{\sc iii--viii},
Ne~{\sc iii--x}, Mg~{\sc iii--xii}, Si~{\sc iv--xiv}, S~{\sc iv--xvi}
and Fe~{\sc vi--xxvi}.  Details of the atomic physics used will be
presented in a subsequent paper. Second, the appropriate fraction of
emission-line photons that escape (in narrow profiles) directly from
the gas without undergoing any further continuum absorption or Compton
scattering is kept track of separately from the rest of the diffuse
radiation originating within the gas \citep{ros79}. Finally, the L
shell lines of Fe~\textsc{xvii}--\textsc{xxii} were changed from
3d$\rightarrow$2p transitions to 3s$\rightarrow$2p transitions. Like
the other prominent lines, the Fe L-lines are produced by
recombination.  \citet{lkog90} have pointed out that recombination
cascades favour the production of 3s$\rightarrow$2p lines over
3d$\rightarrow$2p lines, and this effect is enhanced by resonance
trapping of 3d$\rightarrow$2p lines in optically thick gas. We assume that all 
recombinations to Fe~\textsc{xvii-xxii}, except for radiative 
recombinations directly to the 2p level, ultimately 
result in 3s$\rightarrow$2p emission.

\begin{figure}
\centerline{
\includegraphics[angle=-90,width=0.5\textwidth]{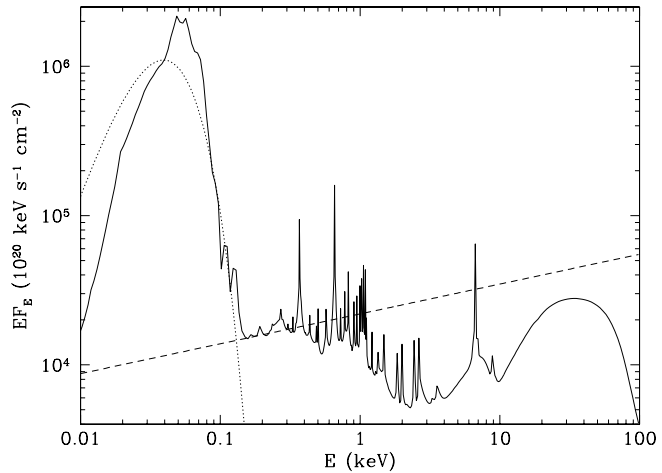}
}
\caption{Computed reflection spectrum (solid line) with $\xi=500$~\ergcms\ and
solar abundances. The incident $\Gamma=1.8$ power-law and 10~eV
blackbody are displayed as dashed and dotted lines, respectively. The
power-law and reflection spectrum were added together prior to the
equivalent width calculations. }
\label{fig:spectrum}
\end{figure}

The results of \citet{ros93} and \citet{ros99} have shown that, over a
wide range of $\xi$, the Ly$\alpha$ line of \oeight\ is typically the
strongest soft X-ray emission line in the reflection
spectrum. Therefore, this is the line that is most likely to be
detected by current instruments and so we computed models where it is
expected the be at its most prominent. Specifically, reflection
spectra were calculated for $\xi=$250, 500 \& 1000~\ergcms. The
incident spectrum was assumed to be a power-law with photon-index
$\Gamma=1.8$, typical of a Seyfert~1 galaxy
\citep[e.g.,][]{nan97}. Furthermore, a 10~eV blackbody illuminated the
base of the slab (at a Thomson depth $\tau_{\mathrm{T}}=10$) with six
times the flux of the power-law. An example of one of the computed
spectra is shown in Figure~\ref{fig:spectrum}.
To investigate the effects of different metal abundances, a set of
models was computed with five times the solar oxygen abundance, and
another set with five times the solar nitrogen abundance. The initial
set of abundances was taken from the list of \citet{mcm83}.

The hydrostatic model requires many input parameters as the height of
the disc at the irradiated point has to be explicitly calculated. For
generality, we chose values typical of a Seyfert~1 galaxy: black hole
mass $M=10^7$~M$_{\odot}$, accretion rate $\dot m =0.01$ (where $\dot
m \equiv \dot M / \dot M_{\mathrm{Edd}}$ and $\dot
M_{\mathrm{Edd}}=48\pi M m_{\mathrm{p}}G/\sigma_{\mathrm{T}}c$, is the
Eddington accretion rate), and the reflection occurs at a disc radius
of 7 Schwarzschild radii. Radiation-pressure dominated boundary
conditions were assumed. As with the constant density models, the
surface of the disc was illuminated with a $\Gamma=1.8$ power-law, but
with an incident-flux-to-disc-flux ratio
$F_{\mathrm{X}}/F_{\mathrm{disc}}=1$. These parameters resulted in a
reflection spectrum similar to the $\xi=500$~\ergcms\ constant density
model (see Sect~\ref{sub:hydro}) and thus isolates the differences due
to the different density structures\footnote{If
$F_{\mathrm{X}}/F_{\mathrm{disc}}=1/6$ rather than 1 for
the hydrostatic model, the added Compton cooling decreases the Compton
temperature and hence the effective ionization parameter of the
atmosphere. The actual differences between these two reflection spectra
in the energy range of interest are fairly small, with the most
significant change being narrower lines due to the O emission region
lying within 0.5 Thomson depths of the surface.}. Only a single model
with solar abundances was computed with these parameters.

\section{Results}
\label{sect:res}
\subsection{Constant density models}
\label{sub:const}
\begin{figure*}
\begin{minipage}{180mm}
\centerline{
\includegraphics[width=0.85\textwidth]{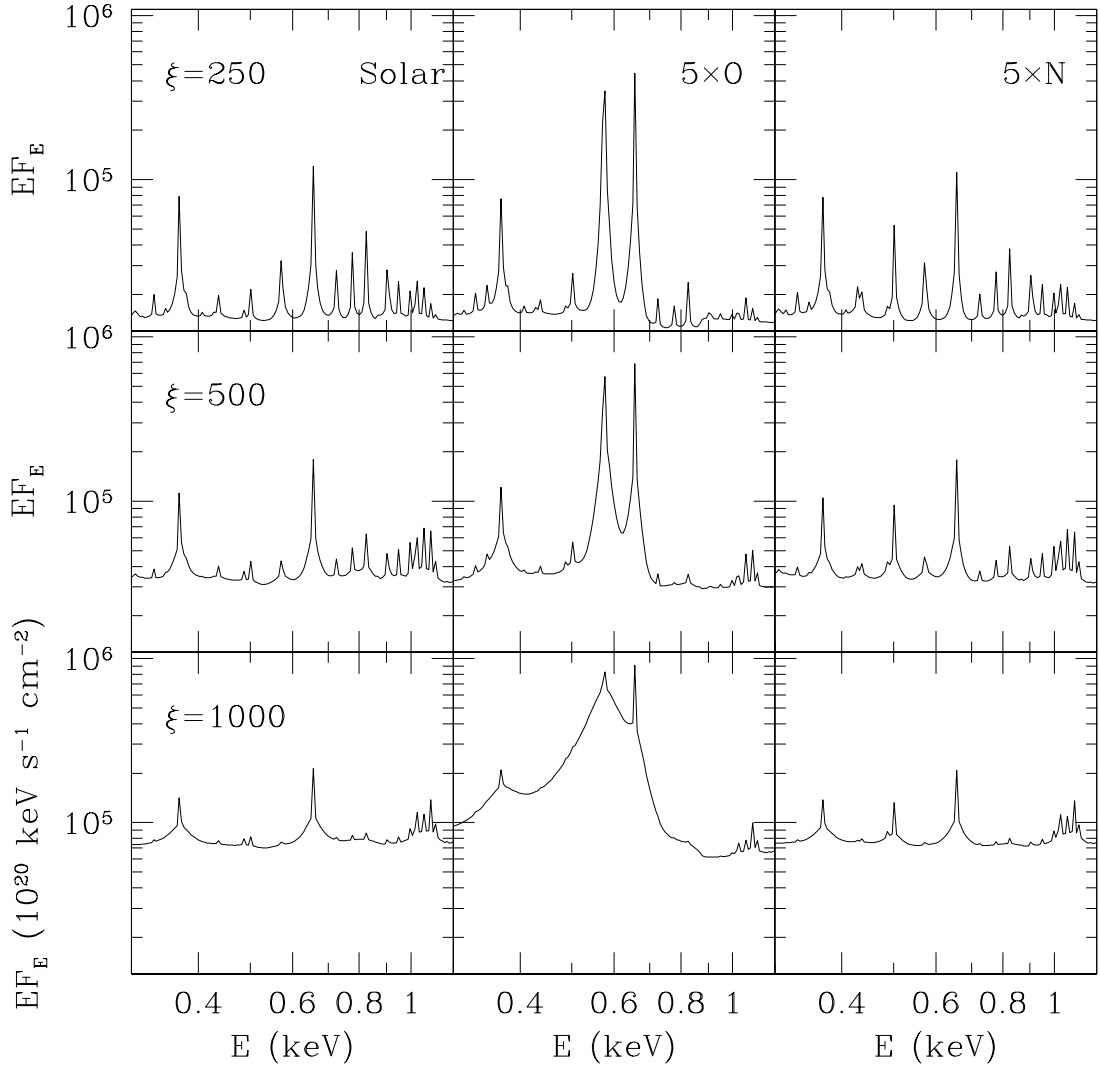}
}
\caption{Total (reflected+incident) spectra between 0.3 and 1.2~\kev\
calculated for $\xi=250$, 500 \& 1000~\ergcms\ and three different
abundance sets (solar, 5 times the solar O abundance, and 5 times the
solar N abundance). Each model included a 10~eV blackbody spectrum
incident on the bottom of the slab with six times the flux of the
irradiating power-law. The most prominent lines present in the spectra are, from left
to right, \csix\ Ly$\alpha$, \nsix\ (resonance+intercombination;
usually very weak), \nsev\ Ly$\alpha$, \osev\
(resonance+intercombination), \oeight\ Ly$\alpha$, and a cluster of Fe
L lines. As the ionization parameter of the slab is increased (moving
downward in the plot), the emission lines weaken significantly
because of the increased ionization level and are broadened due to
Compton scattering. Increasing the oxygen abundance significantly
strengthens the \osev\ line and distorts this part of the spectrum. A
similar, but much less dramatic, effect is seen when the nitrogen abundance
is increased. In this case, the \nsev\ line is most affected,
but the rest of the spectrum is relatively unchanged.}
\label{fig:plot9}
\end{minipage}
\end{figure*}

It is expected that for most Seyfert~1s, the X-ray power-law will be
observed directly along with the reflection spectrum, so our analysis
concentrates on the sum of the incident and reflection spectra (i.e.,
a reflection fraction of one). In Figure~\ref{fig:plot9} we plot the
resulting nine total spectra between 0.3 and 1.2~\kev.
The plots show many lines in this region of the spectrum which vary in
shape and strength as the parameters are changed: \csix\ Ly$\alpha$ at
0.37~\kev, \nsix\ (resonance+intercombination) at 0.43~\kev\ (usually
very weak), \nsev\ Ly$\alpha$ at 0.50~\kev, \osev\
(resonance+intercombination) at 0.57~\kev, \oeight\ Ly$\alpha$ at
0.65~\kev, and a cluster of Fe L lines (particularly from
Fe~\textsc{xvii}--\textsc{xix}) between $\sim$0.7 and 1.0~\kev.  In
Table~\ref{table:data} the equivalent widths (EWs) of the most
prominent lines are listed for each of the nine models, as well as the
width of their blue wings (WBW).
\begin{table*}
\begin{minipage}{190mm}
\caption{Values of the emission line equivalent width (EW) and the
width of the blue wing (WBW) for the spectra shown in
Fig.~\ref{fig:plot9}. Both quantities are tabulated in eV. The WBW was
defined simply as the difference between the base of the blueward wing
and the peak of the line. In general, the EWs decrease as $\xi$ is
increased and the WBW are always on the order of tens of eV.}
\label{table:data}
\begin{tabular}{ccccccccccccccccccc}
 & \multicolumn{6}{c}{Const. density with $\xi=250$~\ergcms} &
 \multicolumn{6}{c}{Const. density with $\xi=500$~\ergcms} &
 \multicolumn{6}{c}{Const. density with $\xi=1000$~\ergcms} \\ 
 & \multicolumn{2}{c}{Solar} & \multicolumn{2}{c}{5$\times$O} &
 \multicolumn{2}{c}{5$\times$N} & \multicolumn{2}{c}{Solar} & \multicolumn{2}{c}{5$\times$O} &
 \multicolumn{2}{c}{5$\times$N} & \multicolumn{2}{c}{Solar} & \multicolumn{2}{c}{5$\times$O} &
 \multicolumn{2}{c}{5$\times$N} \\
Line & EW & WBW & EW & WBW & EW & WBW & EW & WBW & EW & WBW & EW
 & WBW & EW & WBW & EW & WBW & EW & WBW \\ \hline
\csix\ & 26 & 23 & 22 & 27 & 23 & 19 & 19 & 31 & 25 & 47 & 16 & 31
 & 14 & 51 & 13 & 43 & 9.5 & 43 \\
\nsev\ & 2.8 & 10 & 4.1 & 20 & 21 & 36 & 1.8 & 15 & 2.1 & 15 & 18 & 36
 & 0.7 & 10 & 0.2 & 0 & 8.5 & 42 \\
\osev\ & 15 & 35 & 255 & 42 & 14 & 35 & 4.8 & 35 & 163 & 48 & 4.8 & 35
 & 0.3 & 11 & 85 & 67 & 0.2 & 11 \\
\oeight & 73 & 40 & 256 & 40 & 66 & 40 & 51 & 54 & 130 & 62 & 51 & 47
 & 31 & 62 & 19 & 62 & 29 & 62 \\
\end{tabular}
\end{minipage}
\end{table*}
The EWs were calculated by integrating the spectrum over the line
between two hand-picked values of the continuum on either side.  The
WBW is simply the difference between the continuum on the high-energy
side of the line and the value of the spectrum at the line peak, and
is the minimum width of the line. Since the continuum can be difficult
to determine at times (particularly in the cases with high O
abundance), these measurements will not necessarily be very precise,
but will accurately show trends in the data.

Concentrating on the models with solar abundances (the left-hand
column in Fig.~\ref{fig:plot9}), we see that the \oeight\ line is the
most prominent soft X-ray emission feature at all three ionization
parameters with a maximum EW$\approx$73~eV when $\xi=250$~\ergcms. As
$\xi$ is increased to 1000~\ergcms, the N and Fe L lines have all
but disappeared, and the \oeight\ line is drastically weaker
(EW$\sim$31~eV) with broad wings on either side of the line core due
to Compton scattering (WBW$\sim$60~eV).

\begin{figure}
\centerline{
\includegraphics[width=0.5\textwidth]{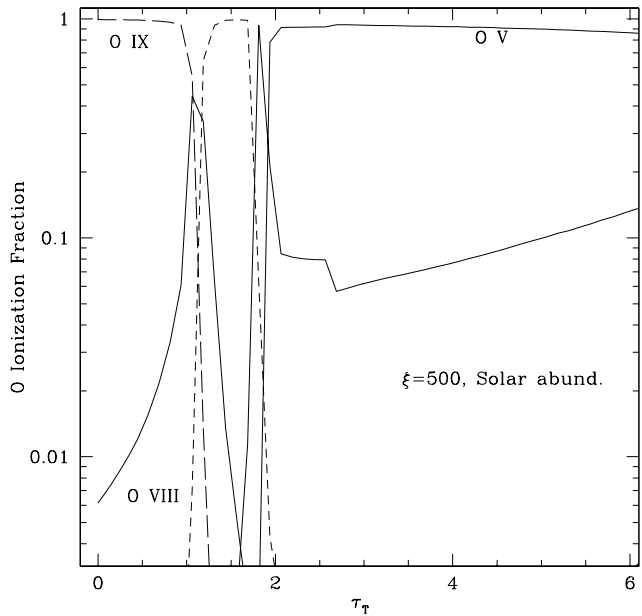}
}
\caption{The oxygen ionization fractions for the $\xi=500$~\ergcms,
solar abundance model. The maximum production of \oeight\ Ly$\alpha$
occurs at a Thomson depth $\sim 1$, so the line photons will
likely be Compton scattered, thereby broadening the profile, as
they pass through the surface of the slab.}
\label{fig:oxy-struct}
\end{figure}

According to Table~\ref{table:data} all of the lines have a WBW on the
order of tens of eV, even at lower ionization parameters. This is
because the majority of the \oeight\ line emission occurs at a
non-negligible Thomson depth and Compton scattering is a major
contributor to the shape of the line. As an illustration, in
Figure~\ref{fig:oxy-struct} the oxygen ionization fractions in the
$\xi=500$~\ergcms\ model are plotted as functions of Thomson depth.
In this case, large fractions of \oeight\ and \osev\ are produced only
at Thomson depths $\ga 1$. Therefore, the line emission
from these species must pass through an optically thick scattering
layer before escaping. As $\xi$ increases the lines originate from a
larger Thomson depth and the photons are scattered more often. 

Increasing the O abundance by a factor of five primarily increases the
strength of the \osev\ line. This is because the incident spectrum is
held constant and the number of possible ionizations from \oeight\ to
O~\textsc{ix} is fixed. With the increase in O abundance, it is no
longer possible to fully-strip a large zone of O, and the curves seen
in Fig.~\ref{fig:oxy-struct} are shifted to the left. Furthermore,
since all of the ionizing photons have been used up, a larger
number of oxygen ions can recombine to the \osev\
configuration. Therefore, the \osev\ line strongly increases with the
oxygen abundance. In fact, at $\xi=250$~\ergcms\ the EWs of both the
\osev\ and \oeight\ lines are $\sim$260~eV which would hopefully be
strong enough to be detected in a high signal-to-noise spectrum of a
Seyfert~1.

In contrast to O, when N is increased by a factor of five, only the
\nsev\ line grows significantly. This is due to the
fact that, even with a five times overabundance, N is not abundant
enough for the \nsev\ edge to saturate in these conditions. The \nsev\
line is strongest in the $\xi=250$~\ergcms\ N-overabundant model, but
still has an EW of only 21~eV. In all other cases, the EW of \nsev\
is $\la 20$~eV. We conclude that at the current time it is unlikely
for \nsev\ Ly$\alpha$ emission to be observed from an irradiated
accretion disc.

The \csix\ line is usually the second strongest line in the reflection
spectrum with EWs between 10 and 30~eV.
In the solar abundance model the \csix\ EW only decreases by about a factor of
two as the ionization parameter is increased from 250~\ergcms\ to
1000~\ergcms. These results certainly suggest that the \csix, \osev\ and
\oeight\ lines are the best candidates for observable soft X-ray accretion disc lines.

\subsection{Hydrostatic model}
\label{sub:hydro}

\begin{figure*}
\begin{minipage}{180mm}
\includegraphics[width=0.5\textwidth]{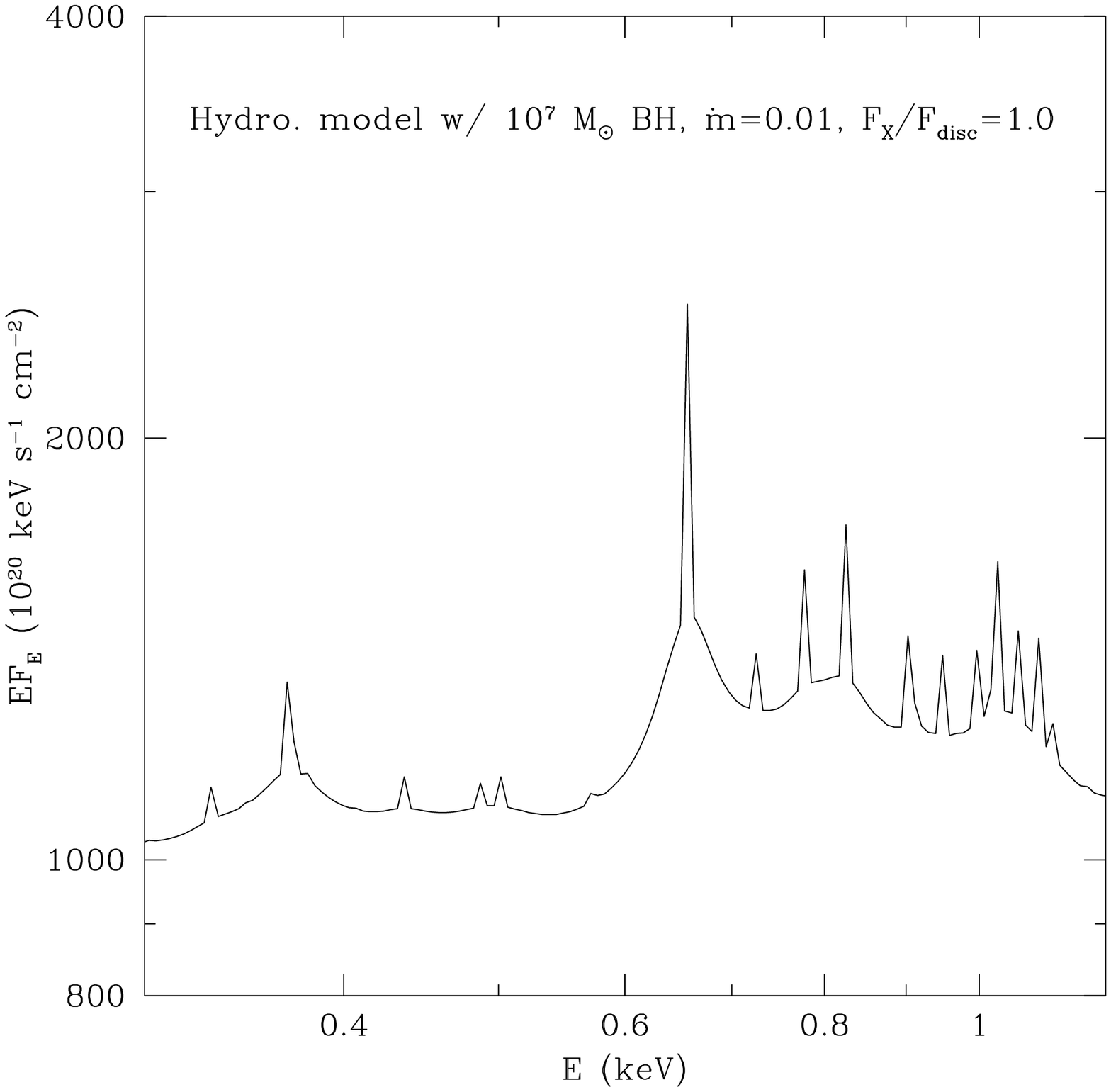}
\includegraphics[width=0.5\textwidth]{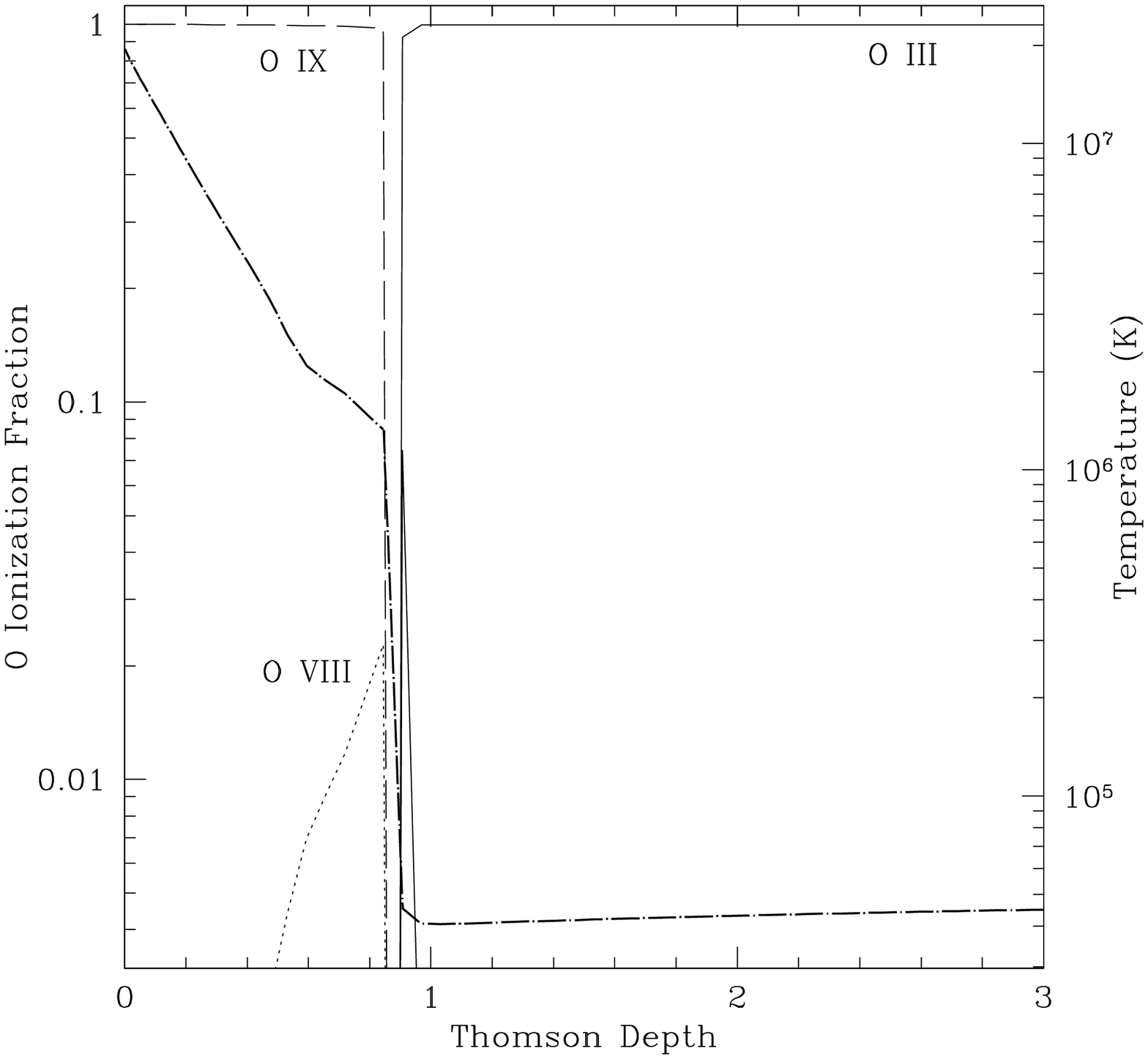}
\caption{(Left) Total (reflected+incident) spectrum between 0.3 and 1.2~\kev\
computed using a hydrostatic ionized disc model. The model assumed a
gas-dominated accretion disc around a 10$^7$~M$_{\odot}$ black hole
accreting at 1 percent of its Eddington rate ($\dot m$=0.01). The
illuminating flux was chosen to be equal to the disc flux. The
spectrum is qualitatively similar to the $\xi=500$~\ergcms\ constant
density model, but the lines are broader due to scattering by the
hot, ionized skin found on the surface of the hydrostatic atmosphere. (Right)
Oxygen ionization fractions for the hydrostatic model shown in the
other panel. The gas temperature is also displayed as the bold
dot-dashed line. The oxygen emission lines mostly originate at a
Thomson depth of $\sim 1$, but must traverse a hot, ionized layer
before escaping. This ionized skin is common in hydrostatic models
and results in broader emission lines than in the analogous constant
density calculation.}
\label{fig:hydro}
\end{minipage}
\end{figure*}

The summed reflection and incident spectrum from the hydrostatic model
is shown in the left panel of Figure~\ref{fig:hydro}.
Qualitatively, the spectrum looks similar to the $\xi=500$~\ergcms\
constant density model with \oeight\ Ly$\alpha$ dominating the
emission. However, many of the lines have broader bases than in the
analogous constant density spectrum. The broad bases are particularly
apparent for the \csix\ (WBW=43~eV), and Fe~L lines.  Furthermore, as
seen in Table~\ref{table:hydrodata}, the lines are actually quite weak
with the EW of \oeight\ only reaching 17~eV. These values are in good
agreement with the results of similar computations done by
\cite{roz02} and Nayakshin (private communication).
\begin{table}
\begin{center}
\caption{As in Table~\ref{table:data}, but using the results of the
hydrostatic model (Fig.~\ref{fig:hydro}). Both quantities are in units
of eV. Although the spectrum is qualitatively similar to the
$\xi=500$~\ergcms\ model, the lines are weaker and broader than those
in the corresponding constant density spectrum.}
\label{table:hydrodata}
\begin{tabular}{ccc}
Line & EW & WBW \\ \hline
\csix\ & 3.1 & 43 \\
\nsev\ & 0.27 & 15 \\
\osev\ & 0.10 & 11 \\
\oeight\ & 17 & 54 \\ 
\end{tabular}
\end{center}
\end{table}

The reason for these differences is that the more realistic density
structure results in a hot scattering layer overlying the line
emitting material. In the right-hand panel of Figure~\ref{fig:hydro}
the vertical temperature profile of the hydrostatic model is overlaid
on a plot of the oxygen ionization structure. We see that the
temperature reaches $kT \sim$1~\kev\ at the surface of the atmosphere,
which is nearly an order of magnitude hotter than the surface of the
$\xi=500$~\ergcms\ constant density model. Oxygen is fully stripped
until a Thomson depth of 0.9, where, because of the rapidly increasing
gas density, it quickly recombines to O~\textsc{iii}, the lowest
ionization state that is treated. Most of the \oeight\ emission
therefore originates from a narrow zone which is not very dense.  Thus
the \oeight\ line is intrinsically weaker in a hydrostatic disc model.
The region in which large fractions of O~\textsc{vii}--\textsc{viii} are
produced is so restricted that the \osev\ line is very much 
weaker in the hydrostatic model than in the constant density
models.  Therefore, the hydrostatic models predict that only 
\oeight\ is detectable.

The existence of the hot, fully ionized skin on the surface of the
disc dilutes the spectral features which are formed underneath by
Compton scattering the emission lines. This results in the broad bases
seen in Figure~\ref{fig:hydro}. Although most of the \oeight\ emission
originates from only a single Thomson depth into the atmosphere, the
temperature of the ionized skin is such to result in significant
dispersion. An ionized skin at the surface of the atmosphere is a
common feature of hydrostatic disc models \citep{nkk00,brf01}. If the
illuminated hydrostatic models are a more realistic description of
accretion discs than the constant density ones, then soft X-ray
accretion disc lines are expected to be weak and broad features, and
will therefore be challenging to detect.

\section{Discussion and Implications}
\label{sect:discuss}

The ultimate goal of X-ray spectroscopy of AGN is to learn about
accretion flows around supermassive black holes. Detecting soft X-ray
emission lines from the accretion disc would complement the
information inferred from the \fe\ line. In principle this should be
relatively straightforward as the effective area of most X-ray
detectors peak in the soft X-ray band, and gratings observations would
allow an accurate determination of any warm absorber. Unfortunately,
the calculations presented in the previous section suggest that most
of the emission lines will be quite weak and broad (even before
relativistic blurring), and therefore difficult to pull out of the
noise with current instruments\footnote{If, as the recent numerical
simulations of \citet{tss02} suggest, the disc is clumpy or
inhomogeneous, then it is much more difficult to predict the properties
of the reflection spectrum.}. The one exception may be the \oeight\
Ly$\alpha$ line which has an EW that can approach 100~eV, depending on
the density structure of the disc. If that line could be
measured accurately enough then, as seen in Fig.~\ref{fig:plot9}, it
would give a good constraint on the ionization parameter of the
disc. If significant \osev\ emission is also seen then abundance
information may also be derived. Although the unambiguous detection of
individual lines will likely be a challenge, soft X-ray accretion disc
lines have the potential to be important to the study of AGN.

\subsection{The soft X-ray features of \mcg}
\label{sub:mcg}
Recently, \citet{br01} and \citet{sak02} claimed to have detected
relativistically blurred \csix\ Ly$\alpha$, \nsev\ Ly$\alpha$ and
\oeight\ Ly$\alpha$ lines in \textit{XMM-Newton} RGS observations of
the Seyfert~1 galaxies \mcg\ and Mrk~766. This interpretation was
challenged by \citet{lee01} who favour a dusty warm absorber model to
explain the features seen in their \textit{Chandra} gratings data of
\mcg. The predictions of ionized disc models should be able to provide
a data-independent view of which interpretation is most
probable. \citet{bf01} fitted the ionized disc models
of \citet{ros93} and \citet{brf01} to the \textit{ASCA} data of \mcg,
concentrating on the region around the \fe\ line. The resulting best
fit models predicted \oeight, \osev\ and Fe~L emission with EWs $<$
20~eV, incompatible with the results of \citet{br01}.

The results presented in Sect.~\ref{sect:res} enable us to take an
orthogonal point of view. Ignoring the \fe\ line entirely, we can ask
the question: is there a region in the explored parameter space which
can account for the putative soft X-ray emission lines? The updated analysis of
\mcg\ by \citet{sak02} is very specific: they only see emission lines of
\oeight, \nsev\ and \csix\ with EWs of $162\pm8$~eV, $54.4\pm3.2$~eV and
$24.9\pm2.5$~eV, respectively. From Table~\ref{table:data} it seems that
a constant density model with $\xi=250$~\ergcms\ and an O overabundance
could account for the C and O lines, but none of the models can
account for the strength of the N line, unless there is also a large (i.e.,
greater than 5 times solar) overabundance of N. However, these models
predict significant \osev\ and Fe L emission which is not seen by
\citet{sak02}, although detectable given the limits placed on the
other lines. The only way to remove these features would be to
increase the ionization parameter, but this will weaken and smear out
the \oeight\ line. Of course, it will be even more difficult to obtain
the required line strengths with hydrostatic models because the
emission features are intrinsically weaker.

Our calculations point to another serious problem with the
relativistic emission line interpretation. Namely, the blue wing of
the putative \oeight\ line is very sharp, dropping down at 700~eV to
the continuum in $\Delta E \ll 10$~eV, as judged from the high
resolution \textit{Chandra} data where the drop is $<$3~eV (Fig.~1 in
\citealt{lee01}). The line must therefore have a relative width of
much less than one per cent. Even a single scattering from an electron
at 10$^6$~K will impart a rms width of 2 per cent to the line, rising
to 6 per cent at 10$^7$~K. A sharp observed feature from a line must
therefore be due to the unscattered line core\footnote{Broadening by
electron scattering is relevant to all narrow emission lines produced
by reflection, and could strongly modify predictions, e.g. those of
\citet{lgk01}.}, which usually carries little flux.

\begin{figure}
\centerline{
\includegraphics[angle=-90,width=0.5\textwidth]{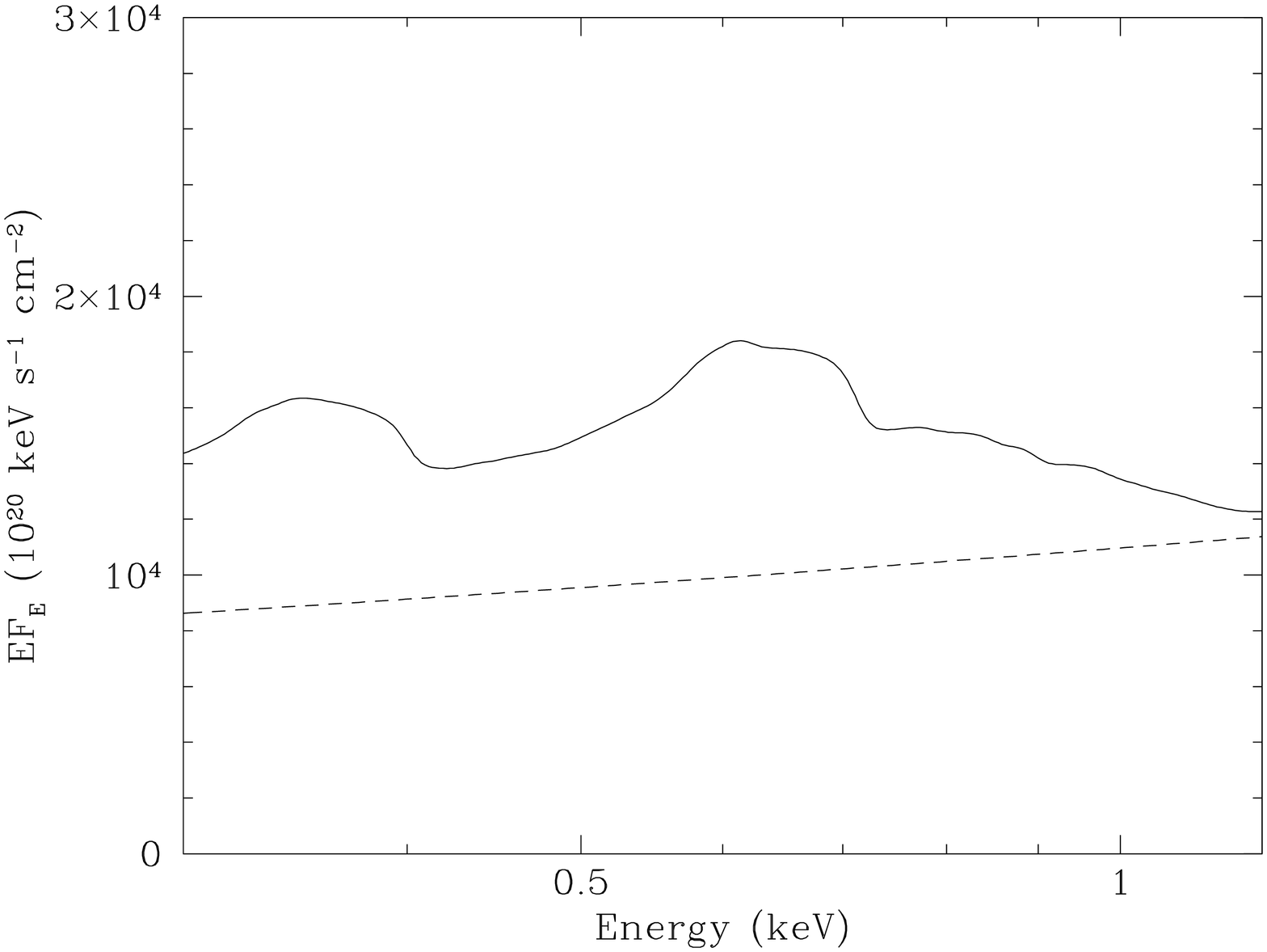}
}
\caption{The $\xi=250$~\ergcms, solar abundance total
(reflected+incident) spectrum between 0.3 and 1.2~\kev\ blurred with
the \mcg\ diskline parameters of \citet{sak02} ($i=38.5$~degrees,
$q=4.49$, $r_{\mathrm{in}}=3.21$~$r_g$, and
$r_{\mathrm{out}}=100$~$r_g$, where $-q$ is the radial emissivity
power-law index, and $r_g=GM/c^2$). The dashed line shows the incident
$\Gamma=1.8$ power-law. The blue wing of the blurred oxygen lines
extends over $\sim$50~eV, as opposed to the $< 3$~eV drop measured by
\textit{Chandra} and \textit{XMM-Newton}. The peak of the line now
occurs at 0.6~\kev.}
\label{fig:blurred}
\end{figure}

The ionized disc models predict a \textit{minimum} width of the
\oeight\ line of over 40~eV (Table~\ref{table:data}). Relativistic blurring
makes the width even larger. Figure~\ref{fig:blurred} shows the
results of blurring the $\xi=250$~\ergcms solar abundance model (the
top-left model from Fig.~\ref{fig:plot9}) with the parameters given by
\citet{sak02}.
With the relativistic blurring effects, the \osev\ and \oeight\ lines
are combined together and peaks at 0.6~\kev. At 0.7~\kev\ the blue
wing of this blended line falls by only 20 per cent over a range of
50~eV. This is significantly weaker than the factor of 2--3 drop over
$\sim$3~eV seen in the \textit{Chandra} and \textit{XMM-Newton}
data. The 0.7~keV drop is weaker in this reflection model because the
Fe~L lines have been blurred together and redshifted against the edge
of the oxygen lines. The oxygen lines can be made sharper by moving to lower
$\xi$, but only at the expense of increasing the strength of the
\osev\ and Fe~L lines. As before, hydrostatic models make the
situation worse because they provide additional Compton scattering due
to the ionized skin. Of course, these considerations also apply to
Mrk~766.

Although we have obviously not covered the entire available parameter
space, we conclude that it is very difficult, if not impossible, for
reflection from ionized discs to produce the lines required by the
interpretation of \citet{br01} and \citet{sak02}. First, there does
not seem to be a situation where one can obtain strong \oeight, \nsev\
and \csix\ Ly$\alpha$ emission without simultaneously having
significant \osev\ and Fe L lines. Second, in any situation where
\oeight\ does dominate the emission spectrum, the disc has to be
fairly highly ionized and so the lines are Compton broadened with
widths of 10s of eV. This width will further increase upon
relativistic blurring, and therefore cannot possibly have the sharp,
$\ll 10$~eV drop at the high-energy side that the \mcg\ data
requires. More realistic hydrostatic ionized disc models exacerbate
these problems because of the hot ionized skin on the surface of the
disc. In general, any soft X-ray relativistic line will tend to create
broad weak features in the observed spectrum. Any large sharp drops in
this spectral range must be due to absorption edges.

It is likely there are weak broad features due to ionized oxygen in
the spectrum of \mcg. They are a plausible identification for the soft
excess required in spectral modelling of \textit{Chandra} data by
\citet{lee01} and for the residuals from a dusty warm absorber fit to
\textit{XMM-Newton} data by \citet[][ see their Fig. 4]{sak02}. On the
basis of our calculations, we conclude that the sharp drop seen in all
soft X-ray spectra of \mcg\ at 0.7~\kev\ cannot plausibly be the blue
wing of an \oeight\ emission line produced by reflection.
 
\section*{Acknowledgements}
The authors thank the referee, Masao Sako, for his helpful comments
and suggestions. DRB acknowledges financial support from the
Commonwealth Scholarship and Fellowship Plan and the Natural Sciences
and Engineering Research Council of Canada. ACF and RRR acknowledge
support from the Royal Society and the College of the Holy Cross,
respectively.


\bsp 

\label{lastpage}

\end{document}